\documentstyle[12pt]{article}
\begin{document}

\begin{center}
{\large GAUGE THEORETIC APPROACH TO FLUID DYNAMICS }\\
\vskip 2cm
Subir Ghosh\\
\vskip 1cm
Physics and Applied Mathematics Unit,\\
Indian Statistical Institute,\\
203 B. T. Road, Calcutta 700035, \\
India.
\end{center}
\vskip 3cm
{\bf Abstract:}\\
The Hamiltonian dynamics of a compressible inviscid fluid  is formulated 
as a gauge theory. The idea of gauge equivalence is exploited
to unify the study of apparantly distinct physical problems and
solutions of new models can be generated from known fluid
velocity profiles.

\vskip 1cm
PACS Numders: 47.10.+g, 03.40Gc, 11.15.-q

\newpage
The notions of gauge symmetry and gauge invariance, or the study of constraint 
systems (in the Hamiltonian framework) 
in a more general sense \cite{di}, have become indispensible tools
in particle physics and to some extent in condensed matter physics. We show in
the present paper that they can play an equally important role in fluid dynamics. 

Formulating a gauge theory to describe the dynamics of a compressible fluid will
be fruitful because a wide class of apparantly distinct 
fluid systems can be shown to be {\it gauge equivalent}, meaning that they will
yield identical results regarding physical quantities in the gauge invariant sector.  
 New fluid models can be generated and solved from the explicit knowledge
of flow patterns of simple systems.  The gauge invariant model that
we construct, following the Batalin-Tyutin prescription \cite{bt}, has the
advantage of having a completely {\it canonical } phase space Poisson Bracket
 structure and thus is eminently suitable for quantization programme and in the
quantum version, powerful results of three or lower spatial dimensional gauge
theories can be borrowed from elsewhere. Note that the construction of the gauge theory
is not unique and alternative gauge models may not have the above attractive features.

It would be interesting to contrast the nature of (gauge) equivalence in
the present theory with 
electrodynamics on one hand and Canonical 
Transfomation (CT) connecting inertial and non-inertial frames on the other
hand. In the former, the Maxwell-Lorentz equations of motion are manifestly
gauge invariant and the gauge transformation does not show up anywhere.
However, in the latter, the same physical situation, if described in
a non-inertial frame, (connected to an inertial one by time dependent
CT), will require inertial or pseudo forces
 involving the transformation parameters. The gauge fluid theory
is a generalization of the latter, since here, (unlike in CT), the Poisson Bracket structure can change under Gauge
Transformations (GT). This and the appearance of interactions arising due to
the GT, (analogous to pseudo interactions in CT), makes the equivalence
non-trivial. 
The compressible fluid equations of motion are 
\begin{equation}
\partial _t \rho +\partial _i(\rho v_i)=X(\rho ,v_j, f(x))~,~~
\partial _t v_i+(v_j\partial _j)v_i=X_i(\rho ,v_j, g(x)),
\label{1}
\end{equation}
where $\rho $ and $v_i$ denote the density and velocity fields
respectively, and $X$ and $X_i$ represent some source and force terms
originating from phenomenological considerations, with $f$ and $g$
being $c$-number functions. $X=0$ signifies
 mass conservation and $X_i=0$ reduces to the "free" Euler equation. The
 aim is to solve these dynamical equations with appropriate boundary
 conditions.

We will demonstrate that (\ref{1}) and another system with the dynamical
variables transformed to $\rho \rightarrow \tilde \rho \equiv \rho ~,~v_i\rightarrow 
\tilde {v_i}\equiv v_i +{\cal A}_i $, are gauge equivalent, provided ${\cal A}_i$
(which can depend on the fields $\rho $ and $v_i$ as well), obeys certain simple
restrictions. Our main result is the explicit form of ${\cal A}_i$.
Thus the theory enjoys a large amount of arbitrariness that 
can be exploited to connect different theories via choice of gauge conditions.
The fact that different sets of source and interaction terms can be traded will have applications both theoretical and experimental studies.

{\it The Hamiltonian system}: Time evolution of fields 
(\ref{1}) are 
 \begin{equation}
 \partial _t\rho (x)=\{H,\rho (x)\}~,\partial _tv_i (x)=\{H,v_i(x)\}~,
 ~~H[\rho ,v_i]=\int dy {\cal H}(y),
  \label{3}
 \end{equation}
with an (in general non-canonical) Poisson bracket structure  satisfying antisymmetry and Jacobi identity.
The free theory corresponding to (\ref{1}) can be derived from the following
Hamiltonian and algebra \cite{th}
 \begin{equation}
 {\cal H}={1\over 2}\rho v_iv_i~,~~\{v_i(x),\rho (y)\}={{\partial }
 \over {\partial x^i}}\delta (x-y)~,~~\{v_i(x),v_j(y)\}=
 -{{\partial _iv_j-\partial _jv_i}\over {\rho }}\delta (x-y).
 \label{4}
 \end{equation}
We consider those systems where additional terms in $H$ can generate
$X$ and $X_i$ in (\ref{1}). For example, a term $p(x)f(\rho (x))$ in 
${\cal H}$ can induce a pressure like term in the $v_i$ equation.
This leaves out viscid fluids, at least for the present work.
 We  use  the 
equivalent Clebsch parametrisation of $v_i$ \cite{th,j}, 
$
 v_i(x)\equiv\partial _i\theta (x) +\alpha (x)\partial _i\beta (x),
 $
satisfying
 \begin{equation}
\{ \theta (x),\alpha (y)\}=-{\alpha \over \rho }\delta (x-y)~,~
 \{\beta (x),\alpha (y)\}=-{1\over\rho }\delta (x-y)~,~
\{ \theta (x),\rho (y)\}=\delta (x-y).
 \label{6}
 \end{equation}

The explicit form of ${\cal A}_i$ is,
\begin{equation}
\tilde {v_i}\equiv v_i+{\cal A}_i=
\partial _i(\theta +\alpha G) +(\alpha +{F\over \rho })\partial _i(\beta -G),
\label{7}
\end{equation}
where different choices of $F$ and $G$, (which can depend on the fields also),
constitute gauge equivalent theories.
The need to restrict $F$ and $G$ is intuitivly clear. The gauge choices, $G=\beta $
 or $F=-\rho \alpha $ will render $v_i$ irrotational, which can not be equivalent
to a velocity field having vorticity, (due to Helmholtz' vorticity theorem). Similarly, if 
invoked, the condition $G
=-\theta / \alpha $ will remove the longitudinal part completely. 
It will be seen that
the above are not proper GTs.

Lastly we mention that since we are concentrating on Hamiltonian systems, the
subsequent gauge equivalent models will also be Hamiltonian, with the following 
structure,
$
\partial _t \tilde\rho =\{\tilde H,\tilde \rho \}_{DB},~
\partial _t \tilde {v_i}=\{\tilde H,\tilde {v_i} \}_{DB},~
\tilde H=H(\tilde {\rho },\tilde {v_i}),
$
where  $DB$ stands for 
(the antisymmetric and Jacobi identity satisfying) Dirac Bracket \cite {di},
 which is a generalization of the Poisson Bracket, appropriate for theories
with constraints. We emphasise  that if the systematic procedure followed here
is not adopted, obtaining the above algebra and 
$\tilde H$ is indeed non-trivial.

{\it Applications}: 
 A crucial feature of our formalism \cite {bt} is the
 generic form 
$$(\tilde \rho ,\tilde v_i)=(\rho ,v_i)
+extension$$
 which ensures
that $\tilde H=H~+~extension$, (if it is polynomial
in nature). Hence the equations of motion in the gauge theory will be of the form,
\begin{equation}
\dot \rho +\partial _i(\rho v_i)=X_0(\rho ,\tilde {v_j}, f)-\partial _i(\rho {\cal A}_i),~~
\dot v_i+(v_j\partial _j)v_i=X_i(\rho ,\tilde {v_j}, g)-\dot {\cal A}_i-(v_j\partial _j){\cal A}_i
-({\cal A}_j\partial _j)v_i.
\label{9a}
\end{equation}
Indeed the above equations are not obvious and are inherent in the BT scheme \cite{bt}.
Notice that compared to (\ref{1}), here we have a different set of 
${\cal A}_i$ dependent source and
interaction terms but the two systems are gauge equivalent because of the 
existence of the gauge invariant set of variables $\tilde\rho \equiv \rho $ and $\tilde {v_i}$,
in terms of which the equations of motion are structurally identical,  \cite{bt}. Experimental or theoretical analysis may become easier in one
system than in the other, such as the point charge-magnetic field 
system simplifies in a suitable rotating frame where the Coriolis
 force removes the magnetic force. (\ref{9a}) is also a new model
which is solved knowing $v_i$, $\rho $ and the chosen form of
 ${\cal A}_i$.

{\it The theory}: It is necessary to
 embed
 the system in an enlarged space having  independent
canonical pairs 
$(\theta ,\Pi _\theta \equiv \rho )$, $(\alpha ,\Pi _\alpha )$ and
  $(\beta ,\Pi _\beta )$ with
  $\{\theta (x),\Pi _\theta (y)\}=\delta (x-y)$ etc. instead of (\ref{6}).
Next we introduce 
 non-commuting Second Class
 Constraints (SCC)  \cite{di} $\eta _a$ which induce
 the previous non-canonical brackets in (\ref{6})
 as Dirac Brackets \cite{di} defined by
\begin{equation}
\{A(x),B(y)\}_{DB}= \{A(x),B(y)\}-\int (d^3zd^3w)\{A(x),\eta _a(z)\}
\{\eta _a(z),\eta _b(w)\}^{-1}\{\eta _b(w),B(y)\}.
 \label{9}
 \end{equation}
The DB
vanishes if either
$A$ or $B$ is $\eta _a $. 
The following SCCs
 \begin{equation}
\eta _1\equiv \alpha \Pi_\theta -\Pi _\beta ~;~~\eta _2\equiv \Pi _\alpha 
 \label{10}
 \end{equation}
reproduce (\ref{6}) as DBs from the canonical set. The
degrees of freedom count remains the same since there are two SCCs
 for the additional variables $\Pi _\alpha $ and $\Pi _\beta $.

Finally we bring in additional
 Batalin-Tyutin (BT) auxiliary variables \cite{bt}
 such that in the final BT-extended phase space the theory is converted
 into a gauge theory, meaning that the final theory has only commuting or
 First Class Constraints (FCC) \cite{di},
  \begin{equation}
 \tilde \eta _1\equiv \eta _1+\phi _1~;~~\tilde \eta _2\equiv \eta _2
 -\Pi _\theta \phi _2~,~\{\tilde {\eta _1},\tilde {\eta _2}\}=0. 
 \label{11}
 \end{equation}
 The BT fields obey $\{\phi _1(x),\phi _2(y)\}=\delta (x-y)$.
  To ensure that there are no further constraints,
  we need a Hamiltonian that commutes with the FCCs. 
The following variables, 
\begin{equation}
\tilde \theta =\theta +\alpha \phi _2 ~,~\tilde \Pi _\theta =\Pi _\theta ~
,~\tilde \alpha =\alpha +{{\phi _1}\over {\Pi _\theta }} ~,~
\tilde \Pi _\alpha = \Pi _\alpha - \Pi _\theta \phi _2 ~,~\tilde \beta =\beta -\phi _2 ~,
~\tilde \Pi _\beta =\Pi _\beta ~,~\tilde \phi _i =0 .
 \label{12}
 \end{equation}
are gauge
  invariant \cite{bt} in the sense that they commute with the FCCs.
   Hence {\it all} quantities written in terms of the redefined variables,
  {\it e.g.} the following (free) Hamiltonian, are gauge invariant in the extended
  space,
   \begin{equation}
 \tilde {\cal H}\mid _{free}={1\over 2}(\tilde \Pi _\theta \tilde {v_i}\tilde {v_i})
 ={1\over 2}\Pi _\theta [\partial _i(\theta +\alpha \phi _2)
 +(\alpha +{{\phi _1}\over {\Pi _\theta }})\partial _i(\beta -\phi _2)]^2.
 \label{13}
 \end{equation}
The remaining interaction terms in $H$ will also be extended in a similar way.
 This Hamiltonian (\ref {13}) together with the FCCs (\ref{11})
and the canonical phase space is the
 gauge invariant system we were looking for.

We note a fortuitous simplification in the extension structures
 (\ref{12}). Unlike in other theories 
 with {\it non-linear } SCCs 
\cite{sg}, where some of the extensions turn
out to be infinite sequences of higher order terms in $\phi _i$-s,
the present theory with non-linear SCCs (\ref{10}),
 is free from this pathology.

To make contact with the physical system, the dimension of the BT extended 
phase space has to be reduced
by additional gauge fixing constraints, (two in this case,
 $\tilde {\eta _3}$ and $\tilde {\eta _4}$,
 corresponding to two FCCs), with the
 only restriction that $\tilde {\eta _a},~a=1,..,4$ constitute an
 SCC system that is $\{\tilde \eta _a,\tilde \eta _b\}\ne 0$. A
 consistency check is to see that
 the original system is recovered in 
 the so called unitary gauge, $\tilde \eta _3
 =\tilde \eta _4 =0$. 
(Also it is now clear that some of the unphysical gauge choices mentioned before are
not valid ones.)
For a particular gauge, one has
 to construct the corresponding $DB$ and compute the equations
 of motion using the $DB$s in reduced phase space, where the SCCs have
been used strongly. Once again, the degrees of freedom count agrees
with the original one.
Consider the special class of gauge transformations: $\phi _1=0~,~
\phi _2=constant $. These will {\it not} change the $(v_i,\rho )$
 algebra. Hence they can be identified as the conventional canonical
  transformations. 
It might be convenient, (although not necessary), to consider the gauges
of the form
 $ \tilde {\eta _3}\equiv \phi _1-F~,~\tilde {\eta _4}\equiv \phi _2
-G$, to remove  the BT fields directly to obtain
(\ref{7}). 
 
Furthermore, additional constraints, such as incompressibility
\cite{t}, can
be included in this setup in the form $\rho =constant$, which under
time persistence generates another constraint $\partial _i(\rho \tilde
{v_i})$. This SCC pair leads to \cite{t}.

The constants of motion for the free theory are obviously the energy
$\tilde H$, the momenta $\tilde P_i=\int (\rho \partial _i\theta +
\Pi _\alpha \partial _i\alpha + \Pi _\beta \partial _i\beta +
\phi _2 \partial _i\phi _1 )$, the angular momenta
$\tilde L^{ij}=\int (r^i\tilde {\cal P}^j-r^j\tilde {\cal P}^i)$ and the boost generator
$\tilde B^i=t\tilde P_i-\int (r_i\rho )$, effecting the
transformation
$$\{\tilde {v_i},u_j\tilde B_j\}=-t(u_j\partial _j)\tilde v_i+u_i~,~
\{\rho,u_j\tilde B_j\}=-t(u_j\partial _j)\rho . $$

Obtaining the Lagrangian is indeed straightforward. The first order
 form is
 $${\cal L}=\Pi _\theta \dot \theta +\Pi _\alpha \dot \alpha
+\Pi _\beta \dot \beta  +\phi _2 \dot \phi _1 -\tilde {\cal H }
-\lambda _1\tilde {\eta _1}  -\lambda _2\tilde {\eta _2}
$$
\begin{equation}
\equiv \Pi _\theta \dot \theta  +\phi _2 \dot \phi _1  +\dot \beta
(\alpha \Pi_\theta +\phi _1)+\dot \alpha \Pi_\theta \phi _2
-\tilde {\cal H},
\label{l}
\end{equation}
where $\lambda _1$ and $\lambda _2 $ are multiplier fields and some
of the variables have been removed using the equations of motion.
At this stage, one can check explicitly that (\ref{l}) is invariant
under the following two independent sets of gauge transformations:
$$
\tilde {\eta _1}\rightarrow \delta _1\Pi _\theta =0,\delta _1\theta =
-\alpha \psi _1,\delta _1 \beta =\psi _1,\delta _1 \alpha =0,
\delta _1 \phi _1
=0,\delta _1 \phi _2=\psi _1;$$
\begin{equation}
\tilde {\eta _2}\rightarrow \delta _2\Pi _\theta =0,\delta _2\theta =
\phi _2 \psi _2,\delta _2 \beta =0,\delta _2 \alpha =
-\psi _2,\delta _2 \phi _1
=\Pi _\theta \psi _2,\delta _2 \phi _2=0,
\label{gt}
\end{equation}
where $\psi _1$ and $\psi _2$ are gauge transformation parameter functions.
Naively taking the unitary gauge, {\it i.e.}  $\phi _1=\phi _2 =0$, we can
recover the Lagrangian posited in \cite{j}.

We conclude by mentioning some of the future prospects. From the
fluid mechanics point of view, the major interest lies in finding
explicit forms of the GT leading to realistic situations. It will also
be interesting if the BT fields can be identified with other
physical variables such as temperature, pressure etc. Notice the
coincedence that the fluid
 gauge theory and the Maxwell electrodynamics have the same 
number of fields and FCCs. Hence it might be worthwhile to look
for some non-linear transformations connecting the two \cite{h}. 
Finally, studying symmetry properties of the gauge fluid theory 
along the lines of \cite{j} can be rewarding.

Acknowledgements: I thank Dr. K. Kumar for numerous stimulating discussions at 
various stages of the work.

\end{document}